# On a new type of non-stationary helical flows for incompressible couple stress fluid

**Sergey V. Ershkov [1],*, Evgeniy Yu. Prosviryakov [2,3], Mikhail A. Artemov [4] and Dmytro D. Leshchenko [5]**

[1] Plekhanov Russian University of Economics, Scopus number 60030998, 36 Stremyanny Lane, Moscow, 117997, Russia; sergej-ershkov@yandex.ru
[2] Institute of Engineering Science UB RAS; evgen_pros@mail.ru
[3] Ural Federal University, Ekaterinburg, Russian Federation; evgen_pros@mail.ru
[4] Department of Applied Mathematics, Informatics and Mechanics, Voronezh State University, 394018 Voronezh, Russia; artemov_m_a@mail.ru
[5] Odessa State Academy of Civil Engineering and Architecture, Odessa, Ukraine, e-mail: leshchenko_d@ukr.net
* Correspondence: sergej-ershkov@yandex.ru

We have explored here the case of three-dimensional non-stationary flows of *helical type* for the incompressible couple stress fluid with given *Bernoulli*-function in the whole space (the Cauchy problem).
In our presentation, the case of non-stationary *helical* flows with constant coefficient of proportionality $\alpha$ between velocity and the curl field of flow is investigated.
Conditions for the existence of the exact solution for the aforementioned type of flows are obtained, for which non-stationary *helical* flow with invariant *Bernoulli*-function is considered satisfying to Laplace equation. The spatial and time-dependent parts of the pressure field of the fluid flow should be determined via *Bernoulli*-function, if components of the velocity of the flow are already obtained. Analytical and numerical findings have been outlined including outstandung graphical presentations of various types of constructed solution in illuminating dynamical snap-shots which demonstrate developing in time the structural behaviour of topology of the aforepresented solutions.

## 1. Introduction, system of equations (incompressible couple stress fluid).

The description of flows of viscous incompressible fluids [1-40] is based on the integration of the classical Navier-Stokes equations and the incompressibility equation [31-32]. In this case, viscous internal forces are described by the symmetric Cauchy tensor [31-32]. In other words, for a representative volume of the medium, only translational (translational) degrees of freedom of movement are taken into account. For incompressible fluids, the symmetry of the Cauchy stress tensor postulates the proportionality of the force of internal viscous friction to the Laplace operator. The proportionality coefficient is the coefficient of kinematic or dynamic viscosity.

It is known that the study of the properties of solutions to the Navier-Stokes equations is still far from complete [12]. Various hypotheses about the structure of flows are used to prove various theorems and construct classes of exact solutions [1-32]. One of the approaches to study the properties of solutions for the Navier-Stokes equations is the use of a regularizing perturbing force proportional to the linear biharmonic Laplace operator [1]. This mathematical way of integrating the Navier-Stokes equations has a clear physical interpretation. If for a representative volume we take into account not only translational degrees of freedom, but also rotational degrees of freedom, with further the clear defining relations for describing fluid flows which contain the biharmonic Laplace operator [1-4].

For the analytical integration of the Navier-Stokes equations, the class of Beltrami flows is known, for which a significant supply of exact solutions has been constructed [5-10]. These exact solutions are important for understanding the mechanisms of interaction of convective mixing of a liquid with internal friction forces [5-10]. The exact solutions obtained can illustrate the chaos in dynamic systems.

The article [13] presents several classes of exact solutions for the Navier-Stokes equations with paired couple stresses. The results of this article generalize the pioneering exact solutions [1-4] and make it possible to describe spatially inhomogeneous flows of non-classical fluids.

Given the importance of finding classes of exact solutions, this paper studies Beltrami-type flows for flows of non-classical fluids. These solutions will be useful for studying the influence of competing dissipative mechanisms on the structure of the velocity field. In accordance with [1-4], system of equations for incompressible flow of micropolar fluid (or incompressible couple stress fluid flow) with conservative body forces should be presented in the Cartesian coordinates as below, under the no-slip conditions over rigid surface:

$$\nabla \cdot \vec{u} = 0\,, \tag{1}$$

$$\frac{\partial \vec{u}}{\partial t} + (\vec{u} \cdot \nabla)\vec{u} = -\frac{\nabla p}{\rho} + \vec{F} + \nu\, \nabla^2 \vec{u} - \mu \nabla^4 \vec{u}\,, \tag{2}$$

where $\boldsymbol{u}$ is the flow velocity, a vector field; $\rho$ is the fluid density, $p$ is the pressure, $\nu$ is the kinematic viscosity, and $\boldsymbol{F}$ represents external force (*per unit of mass in a volume*) acting on the fluid; notation $\boldsymbol{u}$ or $\vec{u}$ means a vector field.

Besides, we assume here external force $\boldsymbol{F}$ above to be the force, which has a potential $\phi$ represented by $\boldsymbol{F} = -\nabla\ \phi$. As for the domain in which the flow occurs and the boundary conditions, let us consider only the Cauchy problem in the whole space.

Let us search for solutions of the system (1)-(2) in a form of *helical* or *Beltrami* flow [5-11] as below:

$$\vec{\Omega}\,(x, y, z, t) = \alpha(x, y, z)\ \vec{u}(x, y, z, t) \tag{3}$$

here we denote *the curl field* $\boldsymbol{\Omega} = (\nabla\times\boldsymbol{u})$, a pseudovector *time-dependent* field (which means *the vorticity* of the fluid flow); $\alpha$ is variable parameter as in [6] (which differs from the case $\alpha$ = const, considered in [5]).

Using the identity $(\boldsymbol{u}\cdot\nabla)\boldsymbol{u} = (1/2)\nabla(\boldsymbol{u}^2) - \boldsymbol{u}\times(\nabla\times\boldsymbol{u})$, we could present momentum equations (2) for incompressible couple stress fluid flow $\boldsymbol{u} = \{u_1, u_2, u_3\}$ as below [6]:

$$\frac{\partial \vec{u}}{\partial t} = \vec{u}\times\vec{\Omega} + \nu\,\nabla^2\vec{u} - \mu\nabla^4\vec{u} - \left(\frac{1}{2}\nabla(\vec{u}^{\,2}) + \frac{\nabla p}{\rho} + \nabla\phi\right) \qquad (4)$$

where we will choose $\rho = 1$ for simplicity. It is worth noting that the continuity equation (1) should be satisfied automatically if we use presentaion of solution in a form (3) for the case $\alpha$ = const. If $\alpha \neq$ const as formulated in more general case in (3), the continuity equation (1) yields demand as below, using (3):

$$\nabla\cdot\vec{u} = 0 \;\Rightarrow\; \left(\frac{1}{\alpha}\right)\nabla\cdot\vec{\Omega} - \left(\frac{\nabla\alpha}{\alpha^2}\cdot(\alpha\vec{u})\right) = 0 \;\Rightarrow\; \left(\nabla\alpha\cdot\vec{u}\right) = 0 \qquad (5)$$

$$\frac{\partial\alpha}{\partial x}\cdot u_1 + \frac{\partial\alpha}{\partial y}\cdot u_2 + \frac{\partial\alpha}{\partial z}\cdot u_3 = 0 \qquad (6)$$

Let us note that the simple choice $\nabla\alpha = \vec{0}$ in (5)-(6) corresponds to the obvious case $\alpha$ = const, we will investigate such the case here in the current research.

## 2. <u>The solving procedure for time-dependent solution, $\alpha$ = const.</u>

Using (3) and (4), we should note that each equation of the *vector* equation in system (4) could be transformed as below in case $\alpha$ = const

$$\frac{\partial \vec{u}}{\partial t} = \vec{u}\times\vec{\Omega} + \nu\,\nabla^2\vec{u} - \mu\nabla^4\vec{u} - \left(\frac{1}{2}\nabla(\vec{u}^{\,2}) + \nabla p + \nabla\phi\right) \;\Rightarrow$$

$$\frac{\partial \vec{u}}{\partial t} = \vec{u}\times(\alpha\vec{u}) - \nu\,\nabla\times(\alpha\vec{u}) - \mu\nabla\times\nabla\times\nabla\times(\alpha\vec{u}) - \nabla B \;\Rightarrow$$

$$\frac{\partial \vec{u}}{\partial t} = (-\nu\,\alpha^2 - \mu\alpha^4)\,\vec{u} - \nabla B \qquad (7)$$

where *Bernoulli*-function $B$ is given by expression below:

$$B = \frac{1}{2}(\vec{u}^{\,2}) + p + \phi\;. \qquad (8)$$

So, first we obtain from (7), using (1):

$$\Delta B = 0 \tag{9}$$

i.e., *Bernoulli*-function $B$ is a harmonic function; the second

$$\begin{cases} \dfrac{\partial u_1}{\partial t} = (-\nu\,\alpha^2 - \mu\alpha^4)u_1 - \dfrac{\partial B}{\partial x}, \\[2ex] \dfrac{\partial u_2}{\partial t} = (-\nu\,\alpha^2 - \mu\alpha^4)u_2 - \dfrac{\partial B}{\partial y}, \\[2ex] \dfrac{\partial u_3}{\partial t} = (-\nu\,\alpha^2 - \mu\alpha^4)u_3 - \dfrac{\partial B}{\partial z}. \end{cases} \tag{10}$$

It is well-known fact that 3D Laplace equation (9) has a fundamental solution (except simple case $B$ = const):

$$B = B_0\sqrt{{x_0}^2 + {y_0}^2 + {z_0}^2}\left(\frac{1}{\sqrt{x^2 + y^2 + z^2}}\right) \tag{11}$$

where, expressions $\sqrt{{x_0}^2 + {y_0}^2 + {z_0}^2}$, $B_0 = const$ correspond to the initial values of the problem as formulated in (1)-(2).

So, we can ajust or freely choose pressure $p$ in (8) in so way that the obtained with help of solving (10) the profiles of velocity should satisfy in (8) where expression of *Bernoulli*-function $B$ is given by equality (11). In this case, we will obtain a kind of exact solution of a type (3) for equations (1)-(2); moreover, each of equations (9) will turn out to be an ordinary differential equation which has fundamental solution as below:

$$
\left\{
\begin{aligned}
u_1 &= \exp\left(-(\nu\,\alpha^2 + \mu\alpha^4)t\right)\left[\left(-\frac{\partial B}{\partial x}\right)\int\left(\exp(\nu\,\alpha^2 + \mu\alpha^4)t\right)dt + C_1(x,y,z)\right] = \\
&= \frac{\left(-\dfrac{\partial B}{\partial x}\right)}{(\nu\,\alpha^2 + \mu\alpha^4)} + C_1(x,y,z)\exp\left(-(\nu\,\alpha^2 + \mu\alpha^4)t\right), \\
u_2 &= \frac{\left(-\dfrac{\partial B}{\partial y}\right)}{(\nu\,\alpha^2 + \mu\alpha^4)} + C_2(x,y,z)\exp\left(-(\nu\,\alpha^2 + \mu\alpha^4)t\right), \\
u_3 &= \frac{\left(-\dfrac{\partial B}{\partial z}\right)}{(\nu\,\alpha^2 + \mu\alpha^4)} + C_3(x,y,z)\exp\left(-(\nu\,\alpha^2 + \mu\alpha^4)t\right),
\end{aligned}
\right. \qquad (12)
$$

where $\{C_1, C_2, C_3\}$ are three functions which should be chosen according to initial values of the problem as formulated in (1)-(2).

Three aforementioned equalities of the system (12) should determine 3 time-dependent functions $\{u_1, u_2, u_3\}$ in regard to the time $t$, with expression for *Bernoulli*-function $B$ given in (11). Expression for pressure field $p$ should be obtained or expressed via equality (8).

## 3. Final presentation of the solution (the helical flows for incompressible couple stress fluid).

Let us present the non-stationary *solution* $\{p, \boldsymbol{u}\}$ ($\boldsymbol{u} = \{u_1, u_2, u_3\}$) of helical flow of type (3) at $\alpha$ = const for the flows of incompressible couple stress fluid (1)-(2) in its final form:

$$p = B - \left( \frac{1}{2}(\vec{u}^{2}) + \phi \right),$$

$$B = B_0 \sqrt{{x_0}^2 + {y_0}^2 + {z_0}^2} \left( \frac{1}{\sqrt{x^2 + y^2 + z^2}} \right), \quad B_0 = const \,,$$

$$u_1 = \frac{\left( - \dfrac{\partial B}{\partial x} \right)}{(\nu\, \alpha^2 + \mu \alpha^4)} + C_1(x, y, z) \exp\left( -(\nu\, \alpha^2 + \mu \alpha^4) t \right),$$

$$u_2 = \frac{\left( - \dfrac{\partial B}{\partial y} \right)}{(\nu\, \alpha^2 + \mu \alpha^4)} + C_2(x, y, z) \exp\left( -(\nu\, \alpha^2 + \mu \alpha^4) t \right), \qquad (13)$$

$$u_3 = \frac{\left( - \dfrac{\partial B}{\partial z} \right)}{(\nu\, \alpha^2 + \mu \alpha^4)} + C_3(x, y, z) \exp\left( -(\nu\, \alpha^2 + \mu \alpha^4) t \right),$$

where $\phi$ is the potential of external force, acting on a fluid; expressions $\sqrt{{x_0}^2 + {y_0}^2 + {z_0}^2}$ $\{C_1, C_2, C_3\}$ correspond to initial values of the problem as formulated in (1)-(2).

We should note if we assume $B_0 = B_0(t)$ in (11), then (12) (and thus, (13)) can be presented in a more general form

$$\begin{cases} u_1 = \exp\left( -(\nu\, \alpha^2 + \mu \alpha^4) t \right) \left[ -\int \left( \left( \dfrac{\partial B}{\partial x} \right) \exp(\nu\, \alpha^2 + \mu \alpha^4) t \right) dt + C_1(x, y, z) \right], \\ u_2 = \exp\left( -(\nu\, \alpha^2 + \mu \alpha^4) t \right) \left[ -\int \left( \left( \dfrac{\partial B}{\partial y} \right) \exp(\nu\, \alpha^2 + \mu \alpha^4) t \right) dt + C_2(x, y, z) \right], \\ u_3 = \exp\left( -(\nu\, \alpha^2 + \mu \alpha^4) t \right) \left[ -\int \left( \left( \dfrac{\partial B}{\partial z} \right) \exp(\nu\, \alpha^2 + \mu \alpha^4) t \right) dt + C_3(x, y, z) \right]. \end{cases} \qquad (14)$$

## 4. Discussion & Conclusion.

The system of equations of motion for incompressible couple stress fluid has already been investigated in many researches including their numerical and analytical findings [1-4], even for 3D case of *non-stationary* flows of incompressible fluid [13]. However essential deficiency exists as ever in the studies of non-stationary solutions of this type of hydrodynamical equations.

We have explored here the case of non-stationary flows of *helical type* for the incompressible incompressible couple stress fluid with given *Bernoulli*-function (11) in the whole space (the Cauchy problem). In this respect we should refer also to the researches [14]-[16] (with an examples of including the Bernoulli-invariant as a key point in solving procedure).

As for the relevance of this new solution, let us discuss the essential details about the possible physical properties of the aforementioned solution (13).

We have schematically imagined a time-dependence of solution (13) for the components of velocity field at Fig.1 below:

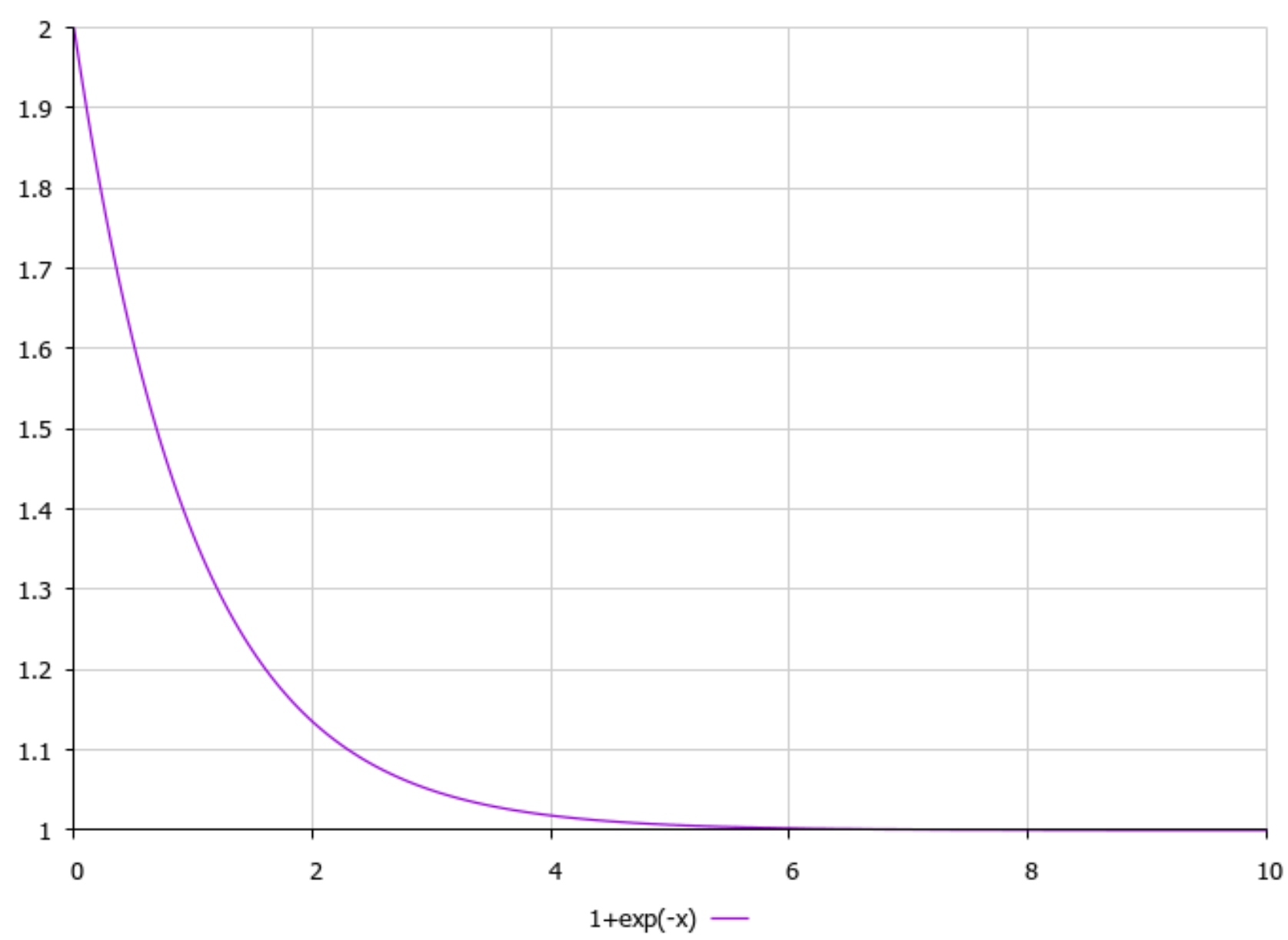

Fig.1. Schematically presented the solution of a type (13) for the components of velocity field (here we designate $x = t$ just for the aim of presenting the plot of solution).

Finalyzing let us outclaim that the case of three-dimensional non-stationary flows of *helical type* for the incompressible couple stress fluid with *Bernoulli*-function satisfying to Laplace equation has been investigated here for *helical* flows taking place in the whole space (the Cauchy problem) with constant coefficient of proportionality α between velocity and the curl field of flow in the current study.

As for conditions for the existence of the exact solution for the aforementioned type of flows, for which non-stationary *helical* flow is determined by given *Bernoulli*-function (here, a fundamental solution of Laplace equation), we can conclude as follows:

1. According to form (3) of *helical* flow strictly, solutions of a type (13) exist if only we choose $B_0 = 0$ or if we choose simplifying condition $B$ = const in (9) from very beginning for the process of constructing the exact solutions; in this case we have from (3)

$$\begin{cases} \alpha C_1(x,y,z) = \dfrac{\partial}{\partial y}(C_3(x,y,z)) - \dfrac{\partial}{\partial z}(C_2(x,y,z)) \ , \\[2ex] \alpha C_2(x,y,z) = \dfrac{\partial}{\partial z}(C_1(x,y,z)) - \dfrac{\partial}{\partial x}(C_3(x,y,z)) \ , \\[2ex] \alpha C_3(x,y,z) = \dfrac{\partial}{\partial x}(C_2(x,y,z)) - \dfrac{\partial}{\partial y}(C_1(x,y,z)) \ , \end{cases} \qquad (15)$$

so, conditions (15) restrict the 9 degrees of freedom in choosing form of functions $\{C_1, C_2, C_3\}$ up to 6 degrees of freedom. Meanwhile, condition $B_0 = 0$ means that we should take into account the potential of external force ϕ (which is for gravity central force known to have a sufficiently large negative value). If we, nevertheless, assume the considering the case of absence of any external force ϕ, this would mean that *helical* flow takes place at negative pressure. As for the negative value of pressure $p$ in (13), we know physically reasonable cases of flows when pressure is transformed to be negative (these are very special conditions for fluids flow, see [39-40]);

2. But nevertheless, it is wortnoting that form (3) of *helical* flow has *already* been taken into account at derivation system of equations (10) (stemming from momentum equations (1)), from which we obtain in result solutions (14) in a most general form. So, using continuity equation (2), we conclude that solutions of a type (14) can exist if restriction to the form of solutions is valid as below

$$\frac{\partial}{\partial x}(C_1(x,y,z)) + \frac{\partial}{\partial y}(C_2(x,y,z)) + \frac{\partial}{\partial z}(C_3(x,y,z)) = 0 \qquad (16)$$

It is also important to note that (16) stems from (15): if we differentiate first equation of system (15) with respect to *x*, second with respect to *y*, third with respect to *z*, then sum them altogether, we should obtain (16) in result. This means that restrictions (15) are more strict and can be considered as excessive with respect to the sufficient conditions (16) for the existence of the exact solution for the aforementioned type of flows.

The spatial and time-dependent parts of the pressure field of the fluid flow should be determined via *Bernoulli*-function, if components of the velocity of the flow are already obtained. Analytical and numerical findings have been outlined including outstandung graphical presentations of various types of constructed solution in illuminating dynamical snap-shots which demonstrate developing in time the structural behaviour of topology of the aforepresented solutions. For example, we can choose in (13) as follows (here below, parameters $B_0 = 0$, $\tau = t$ have been chosen just for simplicity of presentation of velocity's components on Figs.2-5):

$$(\nu\,\alpha^2 + \mu\alpha^4) = 1\,[s^{-1}], \quad \sqrt{{x_0}^2 + {y_0}^2 + {z_0}^2} = 1\ [m^2], \quad B_0 = 0,$$

$$C_1 = \cos((1/\exp\ (\cos(\tau)))\ \cdot 5\tau), \quad C_2 = \sin((1/\exp\ (\cos(\tau)))\ \cdot 5\tau), \quad \tau = t\,, \quad C_3 = 0\,,$$

$$\Rightarrow \quad \begin{cases} u_1 = C_1 \exp\left(-(\nu\,\alpha^2 + \mu\alpha^4)t\right), \quad C_1(x,y,z) = C_1(\tau) \\ u_2 = C_2 \exp\left(-(\nu\,\alpha^2 + \mu\alpha^4)t\right), \quad C_2(x,y,z) = C_2(\tau) \\ u_3 = 0 \end{cases} \qquad (17)$$

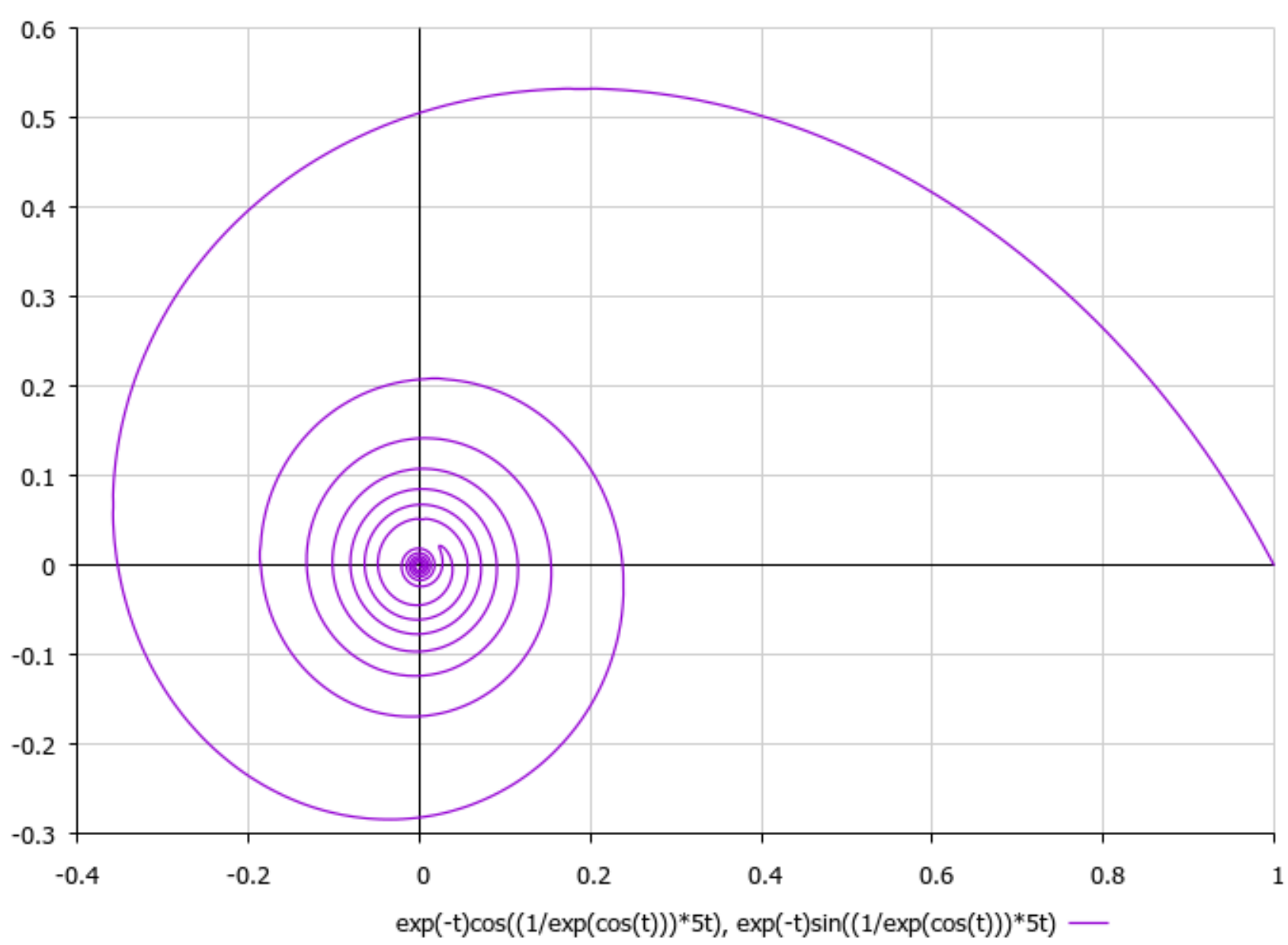

Fig.2. Schematically presented the solution of a type (17) for the components of velocity field (here we designate $x = t$ just for the aim of presenting the plot of solution).

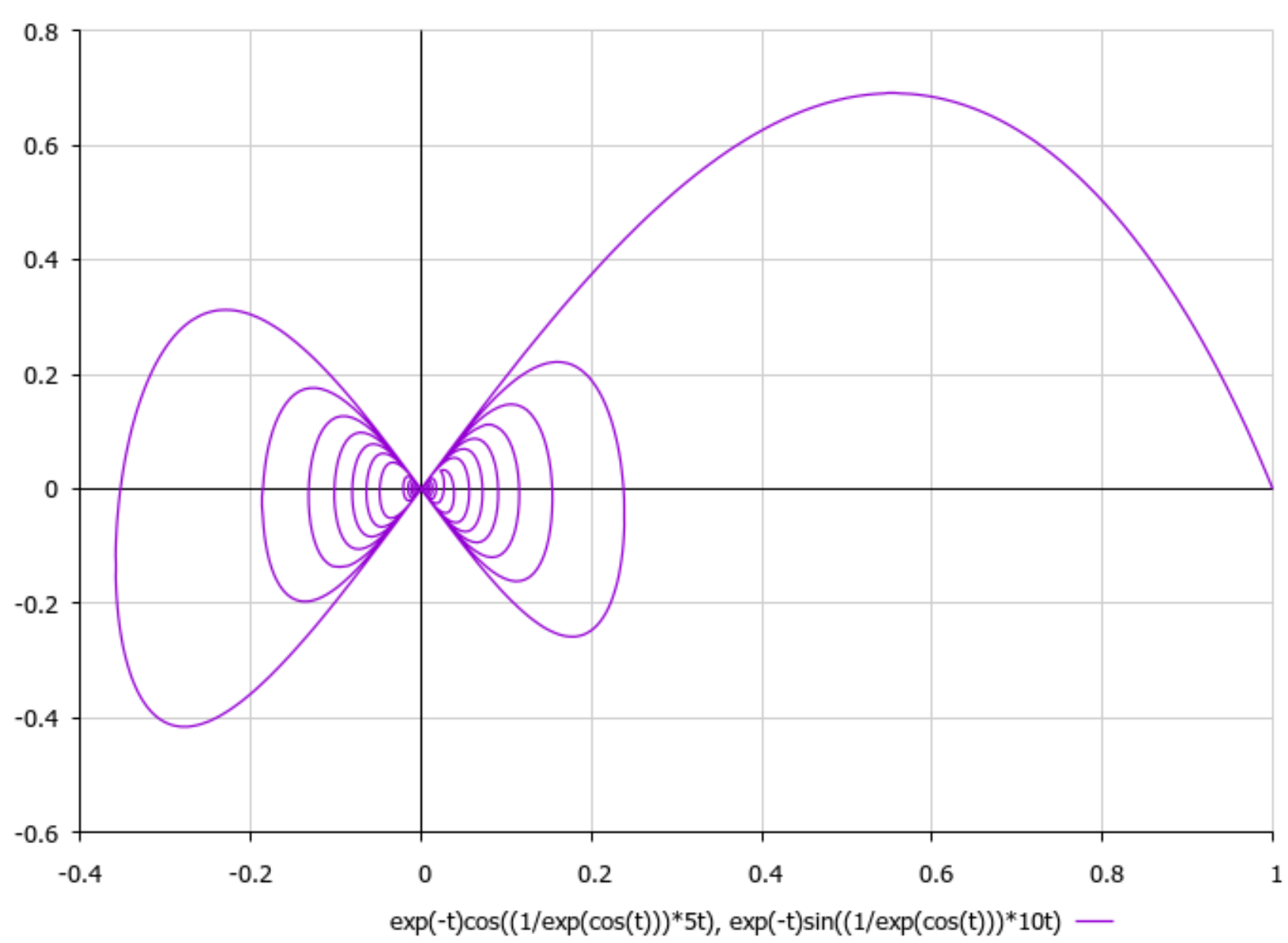

Fig.3. Schematically presented the solution of a type (17) for the components of velocity field (here we designate $x = t$ just for the aim of presenting the plot of solution) if we choose in (15): $C_1 = \cos((1/\exp\ (\cos(\tau)))\ \cdot 5\tau)$, $C_2 = \sin((1/\exp\ (\cos(\tau)))\ \cdot 10\tau)$

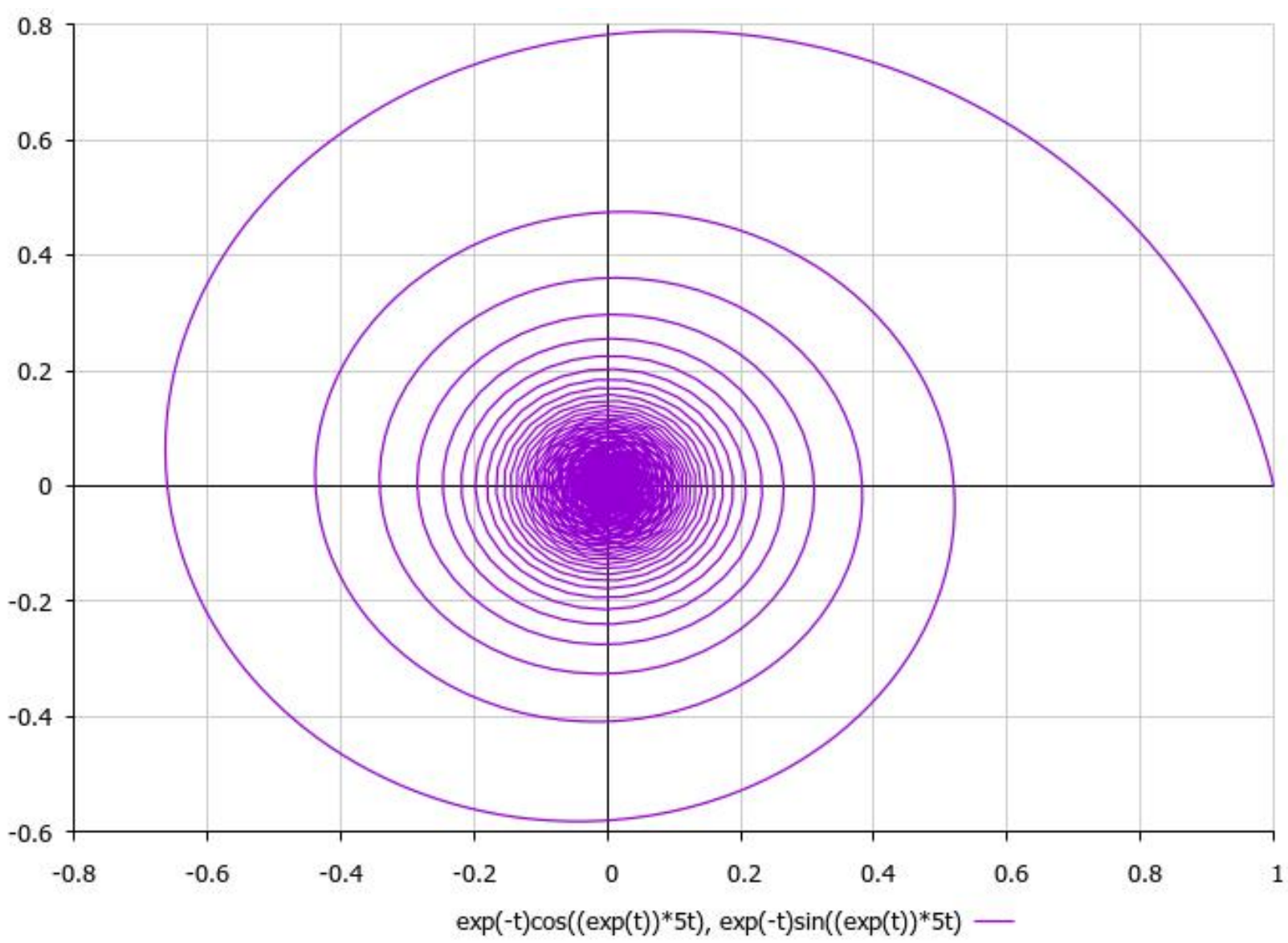

Fig.4. Schematically presented the solution of a type (17) for the components of velocity field (here we designate $x = t$ just for the aim of presenting the plot of solution) if we choose in (15): $C_1 = \cos(\exp(\tau)\cdot 5\tau)$, $C_2 = \sin(\exp(\tau)\cdot 5\tau)$

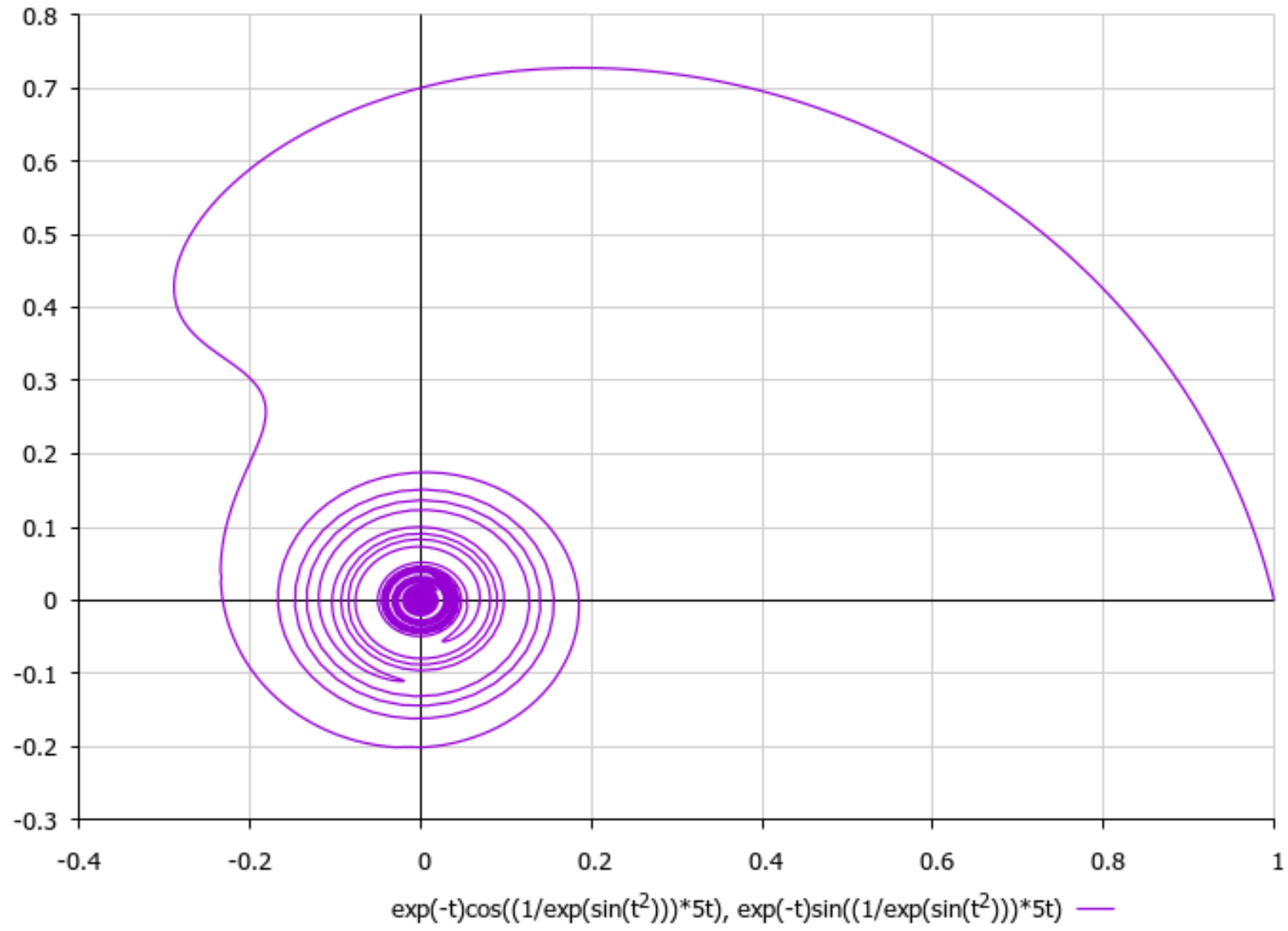

Fig.5. Schematically presented the solution of a type (17) for the components of velocity field (here we designate $x = t$ just for the aim of presenting the plot of solution) if we choose in (15): $C_1 = \cos((1/\exp(\sin(\tau^2)))\cdot 5\tau)$, $C_2 = \sin((1/\exp(\sin(\tau^2)))\cdot 5\tau)$

We should note that in (17) additional restriction should be valid for choosing functions $\{C_1, C_2\}$:

$$\frac{\partial}{\partial x}(C_2(x,y,z)) - \frac{\partial}{\partial y}(C_1(x,y,z)) = 0 \ .$$

While such a theoretical motivation is of course realized, it is worth noting that helical flows is very important in some practical problems, for example in rotor turbine design, fast rotating of coaxial propellers or air-screws of various types of aircrafts, etc.

It should be additionally noted that some mathematical solutions don't reflect physical phenomena and the equation for the force, $\boldsymbol{F}$, used in the analysis, is valid only for conservative forces [6].

The stability of the presented solution is not considered. The last but not least, we should especially mention the comprehensive researches [17]-[25], where a lot of unknown details concerning the close-related area of helical flows are remarked including methods of investigating the stability of such flows.

## **Conflict of interest**

Authors declare that there is no conflict of interests regarding publication of article.

Remark regarding contributions of authors as below:

In this research, Dr. Sergey Ershkov is responsible for the general ansatz and the solving procedure, simple algebra manipulations, calculations, results of the article and also is responsible for the obtaining of exact and analytical solutions.
Prof. Dmytro Leshchenko is responsible for theoretical investigations as well as for the deep survey in literature on the problem under consideration. Prof. Mikhail Artemov is responsible for deep theoretical investigations as well as for theoretical introduction to the problem investigated in the current research.

Dr. Evgeniy Prosviryakov is responsible for obtaining numerical solutions related to approximated ones (including their graphical plots) as well as for theoretical introduction to the problem investigated in the current research.
All authors agreed with results and conclusions of each other in Sections 1-3.